\documentstyle[epsf, epsfig, amssymb, multicol, fullpage]{article}
\begin{document}

\renewcommand{\thefootnote}{\fnsymbol{footnote}}

\title{Olfactory search at high Reynolds number}
\author{Eugene Balkovsky$^1$ and Boris I. Shraiman$^2$ \\
{\small $^1$ James Franck Institute and Department of Mathematics,}\\
{\small University of Chicago, 5640 S. Ellis Ave., Chicago, IL, 60637}\\
{\small $^2$ Bell Labs, Lucent Technologies, 700 Mountain Ave, Murray Hill,
NJ 07974}}
\maketitle

\begin{multicols}{2}

{\bf Locating the source of odor in a turbulent environment --- a
common behavior for living organisms --- is non-trivial because of the
random nature of mixing. Here we analyze the statistical physics
aspects of the problem and propose an efficient strategy for olfactory
search which can work in turbulent plumes. The algorithm combines the
maximum likelihood inference of the source position with an active
search. Our approach provides the theoretical basis for the design of
olfactory robots and the quantitative tools for the analysis of the
observed olfactory search behavior of living creatures (e.g. odor
modulated optomotor anemotaxis of moth).}

Olfactory search strategies are interesting because of their relevance to
animal behavior \cite{PBK,MEC,Malakoff,Barinaga,WZ,Atema} and because of
their potential utility in practical applications (such as searching for
chemical leaks, drugs and explosives \cite{Russel}). Both, the attempt to
characterize and understand the olfactory behavior of living organisms \cite
{PBK,MEC,Malakoff,Barinaga,WZ,Atema} and the more recent effort to design
and build ``sniffing machines'' --- robots that navigate using odors as a
guide \cite{Russel,CGMA,HMG,KGTGL} --- face the common problem of
understanding how the information gained from sporadic detection of odor
dispersed in a naturally turbulent flow can be efficiently utilized for
locating the source. Here we shall discuss the statistical aspects of
turbulent odor dispersal and propose a novel and well defined search
algorithm, which we shall first define in the context of a simplified model
of the turbulent plume and later restate in the form applicable to natural
environment. The proposed search algorithm can be used in robotic
applications and provides a plausible algorithmic interpretation for aspects
of insect olfactory search behavior.

The most familiar strategy for locating the source of a substance is
chemotaxis \cite{Purcell,Berg} which consists of motion in the
direction of the local concentration gradient. Chemotaxis works on
small scales, where the substance spreads by diffusion and the field
of concentration is smooth which is appropriate to the environment of
bacteria, amoebae or single eukaryotic cells \cite{Berg,Devreotes}. On
the other hand, larger animals tracing odors in turbulent flows,
e.g. in atmosphere, have to deal with the complex fluctuating
structure of the odor plume caused by the randomness in the flow which
disperses it. This makes the search a much more complex task.  The
odor is not always present and when it is detected, the local
concentration gradient typically does not point toward the source
\cite {MEC,81MJ,00MWC,TL,SS}. A more complex strategy is required and
additional information such as the current wind direction (and
velocity) is very valuable. One of the best characterized olfactory
search behaviors is the ''odor-modulated optomotor anemotaxis'', which
is employed by males of certain species of moths (see \cite{PBK,MEC})
in order to locate the source of a pheromone (i.e. potential mate),
and which involves, in addition to the sense of smell, the ability to
determine the wind direction\footnote{Moths determine instantaneous
direction of wind visually by looking at the motion of objects on the
ground: when the moth is heading upwind its body axis is aligned with
the direction of motion of the objects (see C. T. David,
Ref.\cite{PBK}, p.  49). }.  The moth's olfactory pursuit flight
exhibits a counterturning pattern: a succession of turns alternatively
to left and right with respect to the wind direction\footnote{It has
been found that counterturns are self-steered as opposed to
gradient-steered \cite{Kennedy,Baker}, which is to say that the moth
makes each turn not because it reaches the boundary of the plume and
feels the lateral gradient of the concentration, but due to internal
turn generator which alternates between right and left.
}. Counterturning is further classified as i) casting and ii)
zigzagging \cite{Kennedy,Baker,DKL,BW96}. The two differ in the extent
of upwind progression: ''zigzagging'' is counterturning with a
significant resultant movement upwind, while ''casting'' is
counterturning with no upwind progression but with wider cross-wind
excursion. Casting and zigzagging behavior has been interpreted
\cite{Kennedy,Baker} as a strategy consisting of upwind progression in
the presence of odor (zigzagging) and cross-wind flight in its absence
(casting). Most generally the two behavioral patterns may be
understood in terms of 'exploitation' and 'exploration'
\cite{behavior1}: the former utilizes the available information, the
latter attempts to collect it. Below we shall identify the
quantitative considerations which provide the rational basis for this
type of behavior in the context of the olfactory search.

Let us first discuss the properties of odor plumes in turbulent flows
\cite {MEC,VILLERMAUX,FR}. If one measures the concentration of odor
far enough downwind of the source, most of the time no odor will be
detected \cite{MEC} . When an odor patch arrives it is detected as a
burst with a complicated small-scale structure, as local concentration
fluctuates strongly while the patch is passing by. The maximal
amplitude of the concentration within such a patch decreases away from
the source, and the average time between two successive patches
increases\footnote{The internal small-scale structure of the odor
burst in principle contains some information about the distance from
the source, but extracting it requires considerably more processing
and we shall limit ourselves to considering the whole burst (or patch)
as a single event.}. The probability of encountering an odor patch
at any given point is determined by the statistics of the flow. The
mean velocity (and direction) of the wind, $V,$ is set by the
large-scale atmospheric conditions and hence stays unchanged for
periods of time long compared with the time-scale of odor
fluctuations. The material points, and thus odor molecules, move with
the local velocity which includes fluctuations about $V$ so that their
net motion is a random walk (due to velocity fluctuations)
superimposed on the drift downwind (due to mean velocity). The
fluctuations of velocity have a correlation length, $L$, which can be
estimated as the height above the ground (since the height above the
ground restricts the size of the vortices) \cite{TL}. At scales larger
than the correlation length $L$, the motion is effectively Brownian
with the diffusion coefficient given by eddy-diffusivity, $D$
\cite{Taylor} which can be estimated as $Lv_{{\rm rms}}$, where
$v_{{\rm rms}}$ is the root-mean-square of velocity
fluctuations\cite{TL}. A patch of odor is blown as a whole along a
Brownian trajectory and is stretched due to spatial inhomogeneity of
the velocity. The stretching produces small-scale variations of the
concentration of odor which decay due to molecular diffusion. Thus,
odor patches have a finite life time due to mixing which occurs
abruptly when the smallest scale of the patch reaches the Batchelor
scale set by molecular diffusion \cite{SS}. Probability of a patch to
survive for time $t$ in the flow is expected to behave as
$e^{-tv_{{\rm rms} }/L}$. Because the fluctuating component of
velocity is typically much smaller than the mean, $v_{rms}<<V,$ the
probability of finding an odor patch at a downstream distance much
larger than $L$ is still significant.

This Lagrangian view \cite{SS} of odor dispersal, in terms of
relatively long-lived patches of odor exercising a biased random work,
suggests the following model which will help us formulate the key
issues relevant to olfactory search. Consider a square two-dimensional
lattice. The sites have coordinates ${\bf r}=(x,y)$ and the source is
at $(0,0)$. At each time step the source releases an ''odor patch''
which is advected by the ''wind''. The wind velocity can take three
values: $(-1,1)$, $(0,1)$, and $(1,1)$, so that the odor patch moves
according to the following rule: each time step its $y$-coordinate
increases by $1$ and $x$-coordinate gets incremented by $\pm 1$ or
stays unchanged. The probabilities of the increments $-1$, $0$, and
$1$ are $p_{{\rm L}}$, $p_{0}$, and $p_{{\rm R}}$.  Without loss of
generality we assume that $p_{{\rm L}}=p_{0}=p_{{\rm R}}=1/3$ , so
that the average velocity points along $y$-axis\footnote{The model
can be readily generalized to include the fluctuations of the
$y$-component of the velocity. One can also include a finite life-time
of particles by introducing the probability for the particle to
disappear.}. This model represents odor dispersion on the length scale
larger than the correlation length, $L,$ which corresponds to the
lattice constant in the model. The three dimensional structure of the
real plume is not essential, because it does not dramatically change
the statistical distribution of 'patches'. On the other hand, it is
natural to conduct a two dimensional search (e.g. constant high above
the ground) at least to within distance $L$ from the source. Hence our
choice of a two-dimensional model.

\begin{figure}[tbp]
\centering\epsfig{file=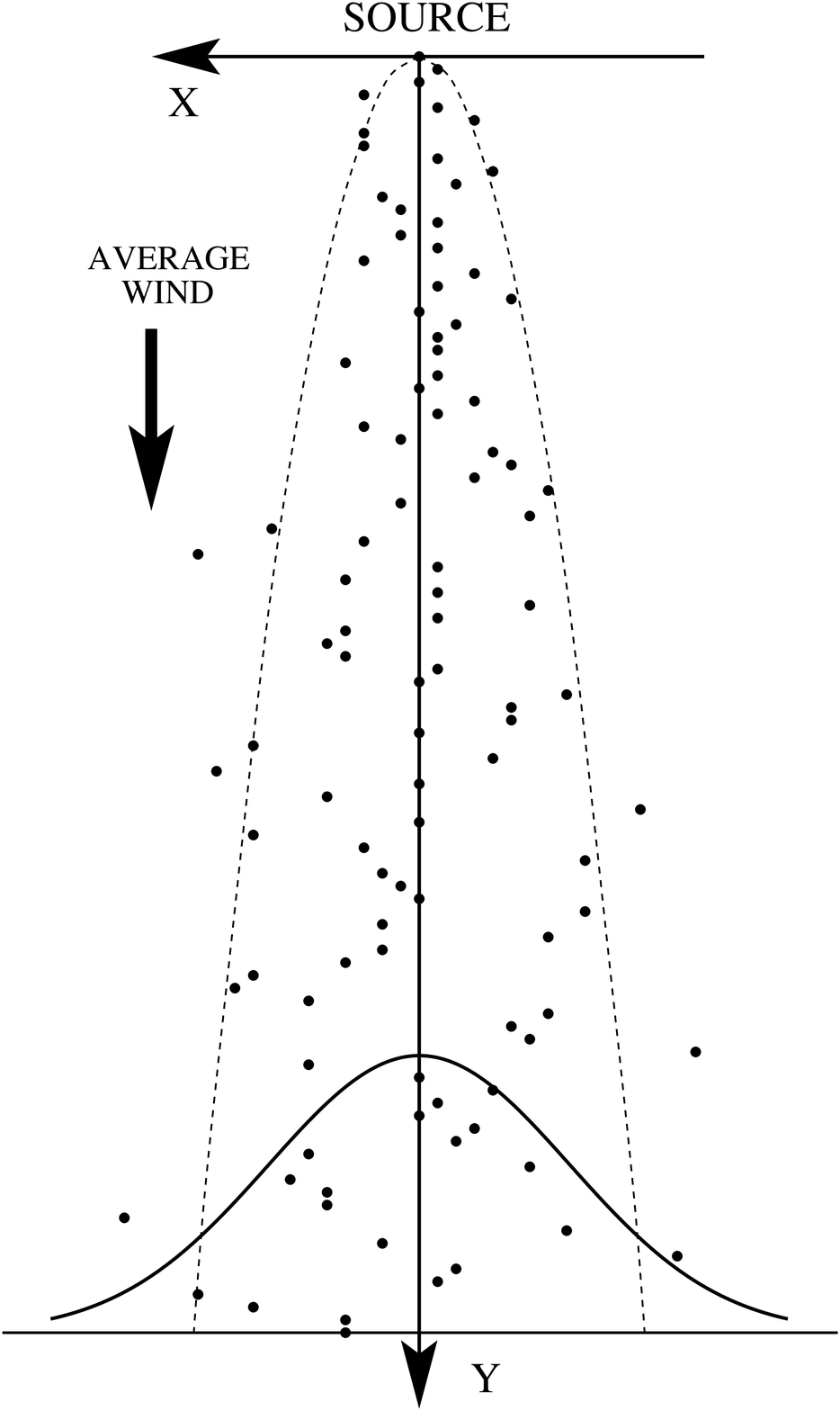,width=9.5cm,angle=0} \vspace{1cm}
\caption{}
\label{fig:plume}
\end{figure}

Figure \ref{fig:plume} shows a snapshot of a plume generated by the process
described above. If we wait long enough, a stationary distribution of
patches will be reached, which at $y\gg 1$ has the asymptotic form 
\begin{equation}
p({\bf r})=\frac{1}{\sqrt{4\pi Dy}}\exp \left[ -\frac{x^{2}}{4Dy}\right]
\,,  \label{gauss}
\end{equation}
where $D=(p_{{\rm R}}+p_{{\rm L}})/2$ is the 'eddy diffusivity'. The
boundary of the time averaged plume has the form $|x|\approx
\sqrt{Dy}$. The probability to find an odor patch at $x\gg \sqrt{Dy}$
is very small. On the other hand, at $x \lesssim\sqrt{Dy}$ the
probability is of the order of $(4\pi Dy)^{-1/2}$.

Let us now define the search rules for the model plume. A local ``observer''
(our model moth or a robot) can detect a) the event of odor patch arrival,
b) direction from which the patch has arrived. This corresponds in reality
to the ability to detect instantaneous odor concentration and instantaneous
wind velocity. Each time step, our robot is able to move at most one lattice
step along $y$-axis and/or one lattice step along $x$ -axis. For simplicity
we will assume that the robot does not move downwind, i.e. it does not
increase its $y$-coordinate. The search starts after the robot contacts an
odor patch for the first time.

Our goal is to find the best search algorithm. An algorithm must determine
where the robot should move at the next moment based on, in principle, all
prior observations. Each algorithm is characterized by the time it takes to
find source --- the search time. Due to random nature of the plume the
search time is a random quantity. Hence we consider the probability
distribution function of the search time, $\rho (t)$, given by the
probability that the source is found during $t,t+1$ interval. For some
algorithms the search time can be infinite in some realizations of the
plume, which means the source is never found. Since each algorithm is
characterized by a function, $\rho (t)$, and not by a number, the definition
of the optimal algorithm is non-unique: one can choose different
optimization criteria. Below we shall seek an efficient algorithm in the
sense of the mean search time.

First consider a simple strategy, which will help to understand the
more efficient one. Suppose that the robot waits at one site until it
gets an odor patch. Then it moves to the site from which the patch
came. This will always lead to the source, i.e. the probability to
miss the source is zero for this strategy. It is possible to
analytically calculate the probability density function (PDF) of the
search time. Near the peak it has the Gaussian form
\begin{eqnarray}&&
\rho (t)\approx \frac{1}{\sqrt{2\pi \Delta }}\exp \left\{ -\frac{(t-t_{{\rm
s }})^{2}}{2\Delta }\right\} \,,\\&& t_{{\rm s}}\propto y_{0}^{3/2}\exp
\left( \frac{x_{0}^{2}}{4Dy_{0}}\right) \,,\,\, \Delta \propto
y_{0}^{2}\exp \left( \frac{x_{0}^{2}}{2Dy_{0}}\right) \,,
\nonumber
\end{eqnarray}
where ${\bf r}_{0}=(x_{0},y_{0})$ is the initial position of the robot, and
$t_{{\rm s}}$ is the 'typical' search time. In addition one can derive that
for $t\gg t_{{\rm s}}$ PDF decays as $\rho (t)\propto t^{-3}$.

From this expression for the PDF one can see that the strategy has
severe drawbacks. If the initial position is inside of the parabolic
region, $x_{0}\lesssim \sqrt{Dy_{0}}$, the typical search time goes
as $y_{0}^{3/2}$.  However, from outside of the parabolic region
$x_{0}\gtrsim \sqrt{Dy_{0}}$, the search time grows with $x_{0}$
faster than exponentially, $\exp \left(
\frac{x_{0}^{2}}{4Dy_{0}}\right) $, which means that the strategy does not
work well for points away from the plume axis. The PDF variance $\Delta $
also increases rapidly with $x_{0}$ and in addition the long time asymptotic
form of the PDF is a power law, $\rho (t)\propto t^{-3}$, which mean that
there exists a relatively high probability that the search takes much longer
than the typical time, $t_{{\rm s}}$. This is explained by the robot's
tendency to get trapped outside of the parabolic region on the way towards
the source.

The same drawbacks are inherent to the maximum likelihood algorithms
\cite{DH,Roz}. In these strategies one estimates the probability for
the source to be located at any given point conditional on the history
of observation and then move in the direction of the most likely
source location. However, unless one waits for a long time so that
many particles are observed, this conditional probability is a flat
function of coordinates, i.e. many source locations around the maximum
have approximately the same probability. As a consequence, all
reliable algorithms of maximum likelihood type suffer from the
drawbacks described above: i) the search time increases rapidly if the
initial position is shifted outside of the parabolic region and ii)
the probability that the search takes much longer than the mean time
is substantial.

\begin{figure}[tbp]
\centering\epsfig{file=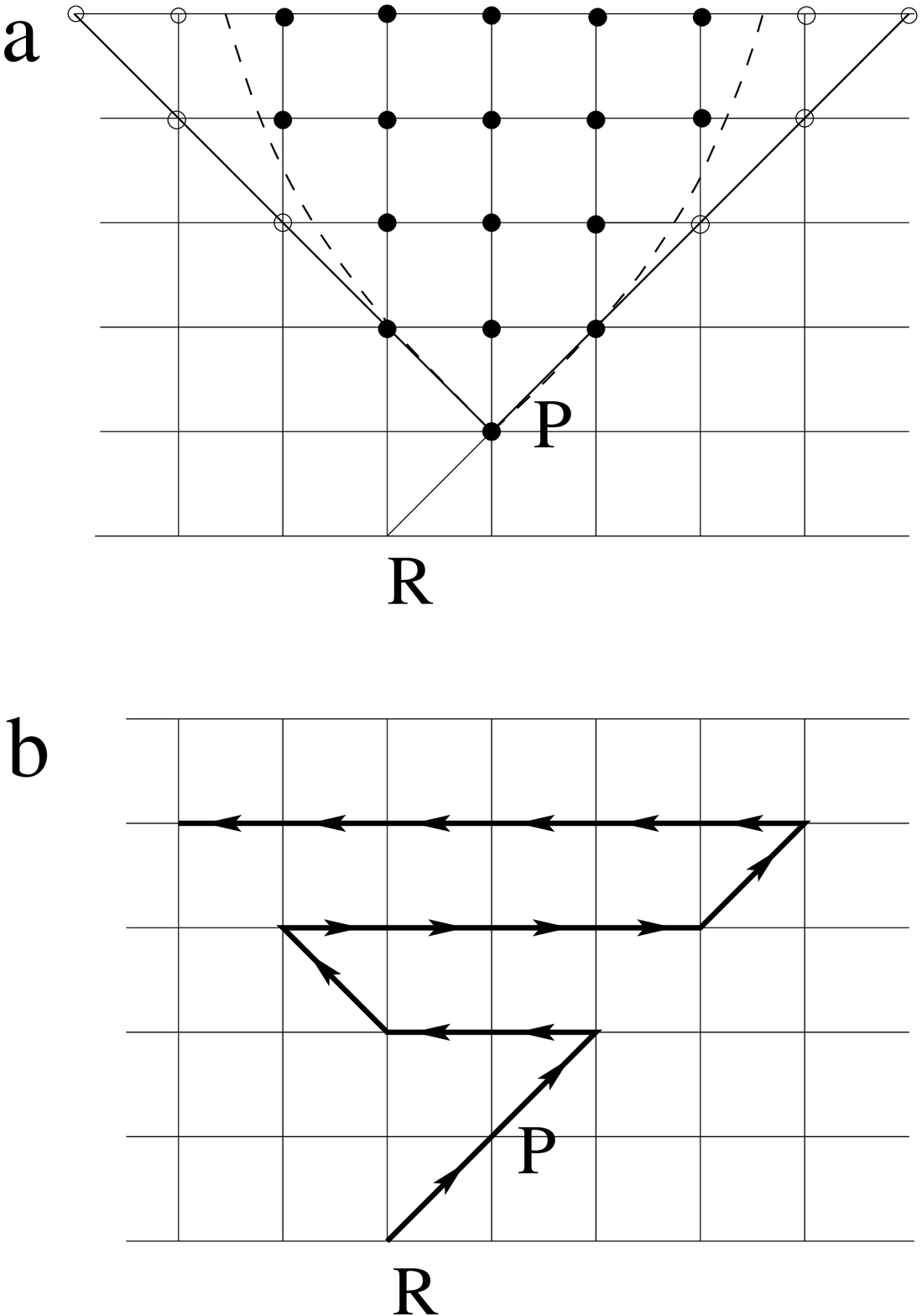,width=9.5cm,angle=0} \vspace{1cm}
\caption{}
\label{fig:cone}
\end{figure}

The inefficiency of the passive search algorithms is related to a
small probability of encountering odor patches. To avoid this, one
should not waste time waiting for odor patches which arrive rarely and
instead actively explore the space. To construct an efficient
algorithm one should take the following facts into consideration. i)
If a patch is observed, it is obvious that the best strategy is to
make a step in the direction from which it has arrived. Each odor
patch observation greatly reduces the uncertainty about the source
position: if $(x_{0},y_{0})$ is the position of the odor patch one
time step ago, the source can only be located in the interior of the
cone $y-y_{0}=\pm (x-x_{0})$, $y<y_{0}$ (see Fig. \ref{fig:cone}). ii)
The probability to encounter a patch at two neighboring sites is
almost equal (see Eq. (\ref{gauss}). iii) In the absence of a patch,
waiting at one site brings very little information about the
source. On the other hand, making one step reduces the uncertainty of
the source location by one point. It follows that making a step is
always preferable over the waiting. When moving, all the points in the
cone must be visited in order not to miss the source. The simplest way
to do this is to visit all the values of $x$ within the cone at given
value of $y$, and then move to $y-1$. In the above example, we visit
the points $(x_{0}\pm 1,y_{0}-1)$ and $(x_{0},y_{0}-1)$ and make sure
that the source is not located (or is located) at one of these
points. After these points have been visited, the closest points which
have to be checked are located at $y=y_{0}-2$. Now there are five of
them: at $x=x_{0}\pm 2$ , $x=x_{0}\pm 1$, and $x=x_{0}$
(Fig. \ref{fig:cone}). This procedure is repeated until the robot
encounters another odor patch. Then the number of possible source
locations gets greatly reduced as the cone of possible positions
collapses to a new one with the vertex at the point of encounter. The
cones are nested, so only the position where last patch was
encountered has to be remembered.

\begin{figure}[tbp]
\centering\epsfig{file=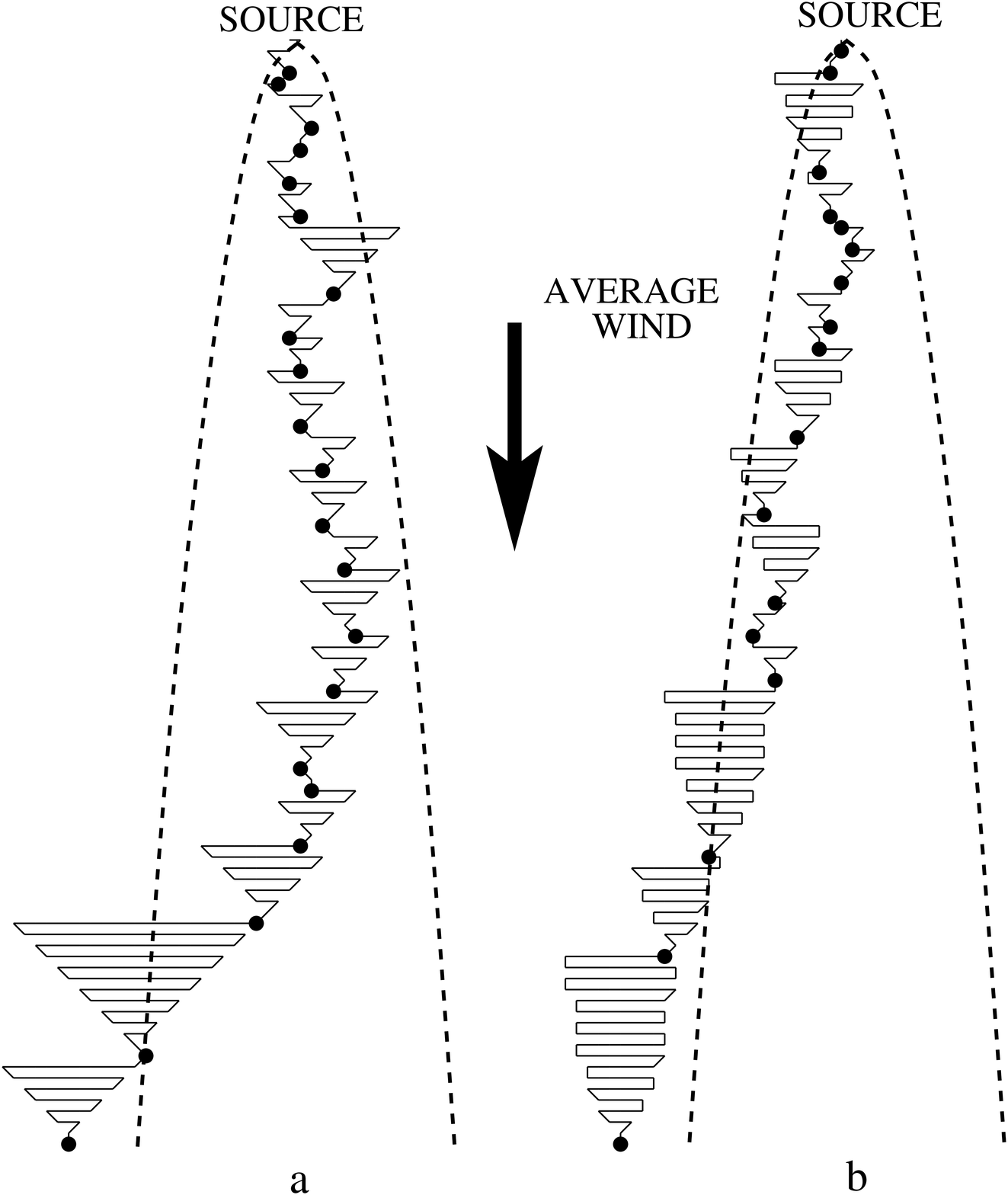,width=9.5cm,angle=0} \vspace{
1cm }
\caption{}
\label{fig:traj1}
\end{figure}

Figure \ref{fig:traj1}a shows a typical trajectory. Since the number
of points to be visited is small after hitting a particle, and grows
thereafter, the amplitude of the 'counterturns' is small immediately
after encounter of an odor patch and then grows. The net upwind
component of the robot velocity is largest after an encounter and gets
smaller with time as $1/\sqrt{t}$.

\begin{figure}[tbp]
\centering\epsfig{file=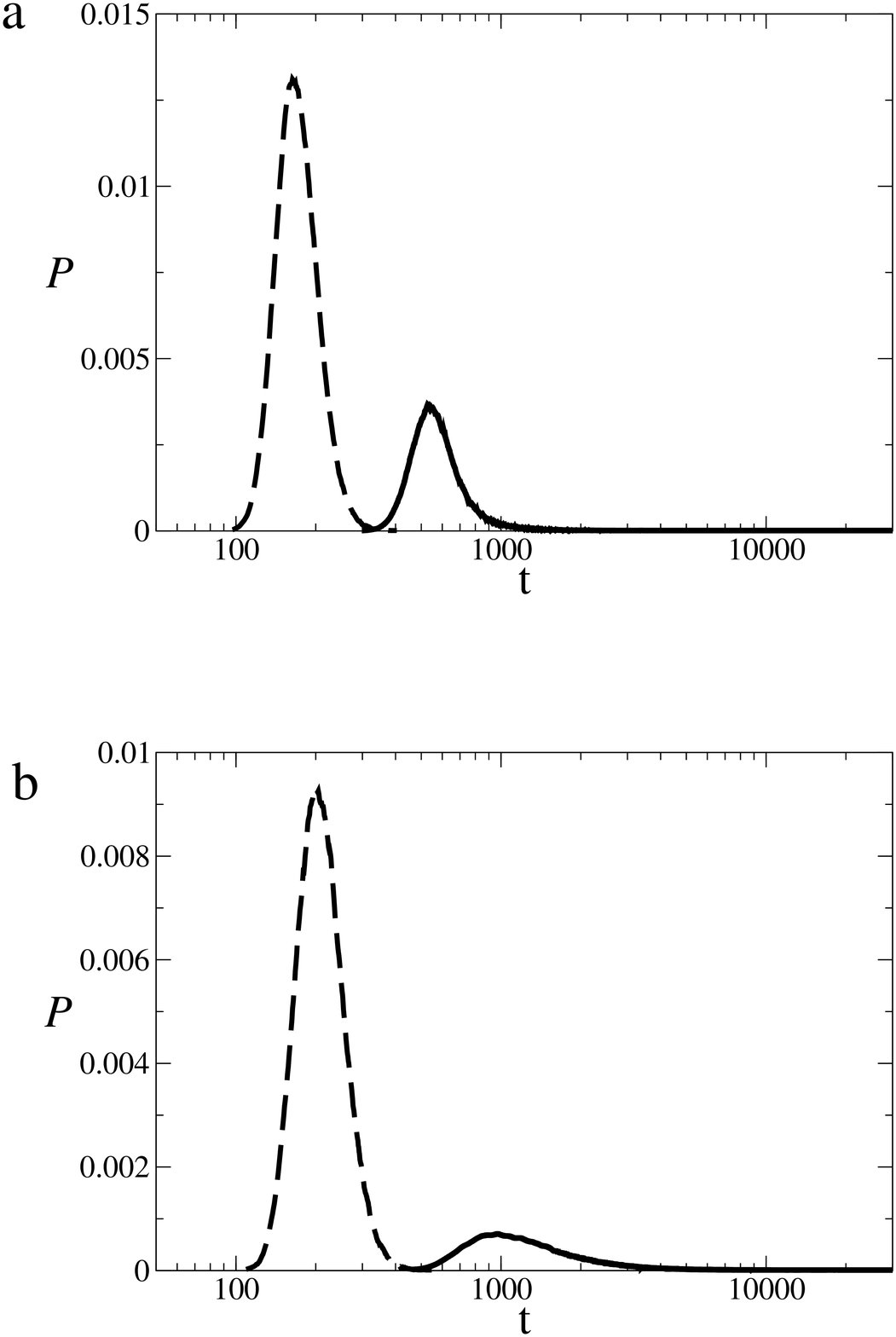,width=9.5cm,angle=0} \vspace{1cm}
\caption{}
\label{fig:hist}
\end{figure}

Within our model one can again analytically calculate the PDF of the search
time 
\begin{equation}
\rho (t)=\frac{1}{4\sqrt{\pi bt}}\exp \left( -\frac{(t-t_{{\rm s}})^{2}}{2bt}
\right) \left( 1+\frac{t_{{\rm s}}}{t}\right) \,.  \label{PDF2}
\end{equation}
with the typical search time 
\begin{equation}
t_{{\rm s}}=ay_{0}^{5/4}.  \label{search_time}
\end{equation}
Here, and $a$, $b$ are constants of the order of one. Most
significantly and in contrast with the ''passive'' strategy considered
before, the search time $t_{s}$ is independent of the crosswind
coordinate $x_{0}$ , which means the search takes approximately the
same time even if the initial position is outside of the parabolic
region. This is a consequence of the ``counterturning'' strategy:
after the first contact with odor, the robot starts to move upwind
with increasing cross-wind amplitude, so with a high probability the
next patch will be encountered inside the region $|x|\lesssim
\sqrt{Dy}$ (see Fig. \ref{fig:traj1}). The PDF has a sharp maximum at
$t\approx t_{{\rm s}}$ and decays exponentially for $t\gg t_{s}$,
i.e. the search time is approximately the same independently of the
realization of the plume. This is explained by the fact that the
number of odor patches encountered by the robot is relatively small,
so most of the time is spent exploring points in the cones. Finally,
the power-law dependence of $t_{s}$ on $y_{0}$ has the exponent $5/4$
which is smaller than the corresponding exponent $3/2$ for the passive
algorithm, so that even the search which starts on the axis of the
plume is faster. This can be seen in Fig. \ref{fig:hist}a which
compares the histograms of the search times obtained by Monte-Carlo
simulations of the two algorithms. The strong dependence of the
''passive'' search time distribution on $x_{0}$ (note the logarithmic
scale of $in$ the figure) is evident even for moderate
$|x_{0}|/\sqrt{Dy_{0}}$ (Fig. \ref{fig:hist}b) - simulating a passive
search which starts further off axis is unreasonably time consuming

Let us now consider a modification of the algorithm, which further
diminishes the search time at the expense of a small probability to lose the
plume. It also allows one to get rid of the lattice and adapt the search
algorithm to a continuous space. In principle, one should visit all the
points in the cone, because the source can be at each point with a non-zero
probability. However, for some points the probability can be quite small.
Such points can be omitted from the search. Let us disregard the points
inside the cone for which the probability to find the source is smaller than
a small constant, $p_{{\rm c}}$. Then from Eq. (\ref{gauss}) we obtain a
parabolic region 
\begin{equation}
(x-x_{i})^{2}\leq 4D(y_{i}-y)\ln \left( \frac{1}{p_{{\rm c}}\sqrt{4\pi
D(y_{i}-y)}}\right) \,, 
\end{equation}
marked by a dashed line in Fig. \ref{fig:cone}. The PDF in this case has the
same form (\ref{PDF2}) with $t_{{\rm s}}=a_{2}y_{0}^{7/6}$. The typical
search time, $t_{{\rm s}}$, has a somewhat weaker dependence on $y_{0}$ than
the other algorithms, however there exists a small probability to miss the
source. A search trajectory for this algorithm is shown on Fig. \ref
{fig:traj1}b.

Although we have formulated the search algorithm in terms of a
two-dimensional lattice model, it is readily generalized to search in
real three-dimensional turbulent plumes. The characteristic
length-scale (the analogue of the lattice constant in the model) is
the correlation length, $L$ set by the height above the ground. The
search is summarized by the following steps: i) detect odor ii) start
crosswind counterturning so that upwind progression per turn is of the
order of $L$ and the transverse amplitude grows as $(Dy/V)^{1/2}$
where $V$ is the mean wind velocity, $D$ is the eddy diffusivity and
$y$ is the upwind displacement from the point of last encounter with
odor. In this way the robot passes within $L$ of any point within the
search domain. The upwind projection of the robot's velocity decreases
as $[v^{2}L^{2}V/(Dt)]^{1/3}$, where $t$ is time elapsed since the
last encounter of odor patch ($v$ is the constant ground speed the
robot). The number of counterturns depends on $t$ as $
[v^{2}t^{2}V/(DL)]^{1/3}$. The essential component of the search
strategy is the crosswind motion, which prevents the searcher from
getting trapped in the regions of exponentially small probability
increasing the rate with which patches of odor are encountered, and
hence increasing the rate of information acquisition. The resultant
transverse motion is biased towards the mid-line of the plume because
that is where odor patches are more frequent. Thus, the algorithm
could be reformulated as the search for the mid-line of the plume
constrained by minimizing the probability of overshooting along the
longitudinal direction.

To conclude, we have proposed an olfactory search algorithm designed
to function in turbulent flow. The efficiency of the proposed
algorithm derives from the implementation of the 'counterturning'
strategy which resembles the observed olfactory search behavior of
moths. The parameters of the counterturns, i.e. amplitude and upwind
drift velocity, are adapted to the measurable statistical properties
of the flow. This algorithm can be readily implemented in a robotic
device, provided the latter is equipped, in addition to a
chemo-sensor, with an anemometer to continuously measure wind
velocity. We have not attempted to rigorously prove the optimality of
the proposed search algorithm. Doing so would be a worthwhile
challenge. Last but not least, our quantitative analysis of search
strategies exposes the role of counterturning and provides insight
into olfactory search behavior of insects and other creatures. It is
likely that zigzagging and casting of the moth are not fundamentally
different but merely correspond to the extremes of a counterturning
behavior \cite{WA91}. Making further comparisons between theoretical
search algorithms and observed search patterns will require new
quantitative experiments with moths in turbulent plumes. The proposed
quantification of the search strategy in terms of PDF of search time
could be applied in experimental context. Another potentially
interesting avenue of research would pursue the connection between the
counterturning search and the very general notion of 'exploration'
versus 'exploitation' in learning behavior\cite{behavior1}.

Authors acknowledge stimulating discussions with A. Gelperin, D. Lee,
and D. Rinberg, and thank A. Gelperin, M. Fee, and H.  Sompolinsky for
careful reading of the manuscript. E.~B.  acknowledges the support by
the National Science Foundation under Grants No. 9971332, 9808595 and
0094569.

\end{multicols}

\newpage \noindent {\bf Figure captions}:

\noindent {\bf Figure 1} A snapshot of the model odor field ($y=100$).
Dashed line bounds the parabolic region where most odor patches are
concentrated. Graph on the bottom represents the probability density
function of patch distribution at $y=100$. Arrow indicates the mean wind
direction.

\noindent {\bf Figure 2} {\bf a}, An odor patch arriving from $
P=(x_{0},y_{0})$ detected at $R$. Circles indicate possible source locations
inside the 'causality' cone with the vertex at $P$. Dashed line is the
boundary of the parabolic region corresponding to relatively high likelihood
of source location. The nodes inside the parabolic region are shown as
filled circles; {\bf b}, A counterturning trajectory inside the cone.

\noindent {\bf Figure3} Typical search trajectories for the two algorithms.
The initial position is $(20,100)$. {\bf a}, The search is performed inside
'causality' cones; {\bf b} The modified search, where only the points inside
the parabolic high-likelihood regions are searched. Arrow indicates the mean
wind direction. The dashed lines show the region of high probability to
encounter odor patch.

\noindent {\bf Figure 4} Histogram of the search time obtained numerically
using Monte-Carlo simulations. {\bf a}, with initial condition $(0,50)$; 
{\bf b}, with initial condition $(10,50)$. Solid line shows the histogram
for the passive search algorithm, dashed line --- for the active search
algorithm. Note the logarithmic scale of $t$.

\newpage


\end{document}